\documentclass[10pt, conference]{IEEEtran}

\IEEEoverridecommandlockouts

\AtBeginDocument{%
  \providecommand\BibTeX{{%
    \normalfont B\kern-0.5em{\scshape i\kern-0.25em b}\kern-0.8em\TeX}}}

\usepackage{graphicx}
\usepackage{xcolor}
\usepackage{hyperref}
\usepackage{url}
\usepackage{cleveref}
\usepackage{xspace}
\usepackage{csvsimple} 
\usepackage{booktabs}
\usepackage{balance}
\usepackage{cite}

\newcommand{\ie}{\emph{i.e.,}\xspace}
\newcommand{\eg}{\emph{e.g.,}\xspace}

\title{Predicting Tags For Programming Tasks \\ by Combining Textual And Source Code Data}

\begin{document}

\author{
 \IEEEauthorblockN{Artyom Lobanov,\IEEEauthorrefmark{1} Egor Bogomolov,\IEEEauthorrefmark{1} Yaroslav Golubev,\IEEEauthorrefmark{1} Mikhail Mirzayanov,\IEEEauthorrefmark{2} Timofey Bryksin\IEEEauthorrefmark{1}}
    \IEEEauthorblockA{\IEEEauthorrefmark{1}\textit{JetBrains Research},  \IEEEauthorrefmark{2}\textit{Codeforces}}
    \IEEEauthorblockA{\{artem.lobanov, egor.bogomolov, yaroslav.golubev,  timofey.bryksin\}@jetbrains.com, mirzayanovmr@gmail.com}
}

\maketitle

\begin{abstract}

Competitive programming remains a very popular activity that combines both software engineering and education. In order to prepare and to practice, contestants use extensive archives of problems from past contents available on various competitive programming platforms. One way to make this process more effective is to provide an automatic tag system for the tasks. Prior works do that by either using the tasks' problem statements or the code of their solutions.
  
In this study, we investigate which information source is more valuable for tag prediction. To answer that question, we compare existing approaches of both types on the same dataset and with the same set of tags. Then, we propose a novel approach, which is an ensemble of the Gated Graph Neural Network model for analyzing solutions and the Bidirectional Encoder Representations from Transformers model for processing statements. Our experiments show that our approach outperforms previously proposed models by $0.175$ of the PR-AUC metric.

\end{abstract}


\section{Introduction}\label{sec:introduction}

\renewcommand*{\thefootnote}{\fnsymbol{footnote}}
\footnotetext[1]{The work and the paper were completed in 2020.}
\renewcommand*{\thefootnote}{\arabic{footnote}}

Competitive programming is a mind sport where contestants are required to write a computer program in one of the supported programming languages to solve an algorithmic problem under certain limitations on time and computational resources. Participation in such competitions increases students' motivation to learn programming, algorithms, and data structures, as well as helps them to pass technical interviews in IT companies~\cite{bloomfield2016programming}.

Competitive programming platforms, such as Codeforces~\cite{codeforces} and CodeChef~\cite{codechef}, have extensive archives of problems from past contests that can be used to prepare for competitions or as a practice for programming courses on algorithms or data structures.
Unfortunately, problems in such archives are often poorly systematized, which makes it hard to find problems for a particular topic. To mitigate this issue, some contest platforms implement tag systems. A tag is a special label assigned to a problem that indicates that it belongs to a certain class. A tag may describe the problem's topic (\eg \textit{math}, \textit{geometry}, \textit{graphs}, \textit{strings}) or possible approaches to solving it (\eg \textit{dynamic programming}, \textit{brute force}, \textit{binary search}). A problem can have multiple tags of both types. 
Tags are often set manually, which takes a lot of time and effort. Moreover, the resulting labeling can be subjective and inconsistent. 
An automated tagging system is intended to solve these problems, and there were several attempts to create such systems~\cite{Bora2016PredictingAA, shrivastava2019predicting, iancu2019multi, shalaby2017automatic, sudha2017classification, intisar2019classification}. 

One possible way to predict tags is to analyze problem statements. 
Since problem statements are usually written in natural languages (\eg English or Chinese), some authors suggested to apply techniques from natural language processing (NLP) to the tag prediction problem~\cite{Bora2016PredictingAA, shrivastava2019predicting, iancu2019multi, intisar2019classification}. These approaches show promising results for tags related to the problem's topic, but determining tags that describe possible approaches to solving the problem based only on its statement is a difficult task, similar to solving the problem itself.

To identify tags that describe how the problem could be solved, we can leverage information from the already submitted solutions to this problem. 
This task requires source code analysis, which is also a rapidly developing area. 
Several prior studies followed this idea and treated tag prediction as an algorithm classification problem~\cite{shalaby2017automatic, sudha2017classification}, which aims to classify snippets of code according to the algorithm they implement. 
Such approaches also have disadvantages. For example, tags are generally used to label problems, not their solutions. Also, some complex problems might have several possible ways to solving them, and as a result some tags may be irrelevant to a particular solution. 

Unfortunately, prior works used different task definitions and evaluation methods, so it is impossible to directly infer which way of predicting tags works better. Since both types of approaches have their advantages and disadvantages, we propose to combine both data sources into a novel technique. The main idea of our approach is to train two separate machine learning models: the first one employs modern NLP methods to analyse problem statements, and the second one extracts the code semantics to analyze the submitted solutions. Then, we combine both models into an ensemble to get more robust results. Since each problem may have several relevant tags, this task constitutes a multi-label classification problem. Our experiments show that this approach outperforms existing solutions that use only one type of input data. 

Our main contributions are: 
\begin{enumerate}
    \item \textbf{Novel approach} for predicting tags in competitive programming problems. In contrast to previous works, our approach leverages information from both problem statements (text) and submitted solutions (source code).
    \item \textbf{Comparison} of existing tag prediction approaches. We reproduced 13 existing approaches from five papers and compared their quality on the same dataset. Our experiments show that our model outperforms the existing approaches by $0.175$ in terms of PR-AUC metric.
    \item \textbf{Replication package} containing our trained model and the source code for all the evaluated models~\cite{rp}. The trained model may be used to predict tags for educational or competitive programming problems.
\end{enumerate} 

The remainder of the paper is organized as follows. We discuss previous works on tag prediction in \Cref{sec:background}. The architecture of our approach is presented in \Cref{sec:approach}. \Cref{sec:dataset} describes the dataset we use to compare the models. Following that, in \Cref{sec:evaluation} we describe the experimental setup and present the results, together with the discussion. The threats to the validity of our research are listed in \Cref{sec:threats-to-validity}. Finally, we conclude our paper in \Cref{sec:conclusion}.

\section{Background}\label{sec:background}

\subsection{Natural Language Processing}

Bora et al.~\cite{Bora2016PredictingAA} applied the Long Short Term Memory (LSTM)~\cite{hochreiter1997long} neural model to predict tags from problem statements. The authors used both pre-trained word2vec~\cite{mikolov2013distributed} vectors and one-hot encoding to represent the input data. 
The authors took statements of problems from Codeforces, a popular competitive programming platform, and only considered the first tag for each problem.
In their experiments, only Random Forest~\cite{ho1995random} (one of the machine learning models they tested) outperformed a dummy classifier, which was used as a baseline and always predicted the most popular class. 

Athavale et al.~\cite{shrivastava2019predicting} used several approaches based on the application of Convolutional Neural Networks (CNN)~\cite{Kim2014ConvolutionalNN} to text analysis. They investigated the tag prediction problem as both multi-class classification (mapping each entry to one of several classes) and multi-label classification (mapping each entry to several classes at once). Again, a dataset of problems from Codeforces was used. The authors tried to predict 10 and 20 most frequent tags, and the best results were obtained by an ensemble of CNNs. The authors also asked several people with some experience in competitive programming to predict tags. Human results turned out to be better than the result of the best model, but still not very good: F1-macro score of 0.43 on the top-20 multi-label classification task. This indicates that assigning tags to problems is a difficult task even for humans.

The same task was investigated by Iancu et al.~\cite{iancu2019multi}. The authors collected problem statements from Codeforces and TopCoder~\cite{topcoder}, and sorted their tags into 9 classes. The authors employed the following approaches: Doc2Vec~\cite{Quoc2014documents}, LSTM over word2vec and one-hot encoding. The best F1 score was achieved by LSTM over one-hot encoding, but LSTM over word2vec was better according to the Weighted Hamming Score (the weighted version of Hamming distance~\cite{hamming1950}). 

Intisar et al.~\cite{intisar2019classification} used a slightly different approach. First, they applied topic modeling algorithms (LDA~\cite{Blei2003LatentDA} and NMF~\cite{Lee2000AlgorithmsFN}) to vectorize text and then used these vectors to train and evaluate several classification algorithms, such as kNN~\cite{larose2014}, Random Forest, Multinomial Naive Bayes (MNB)~\cite{rennie2003tackling}, and Multilayer Perceptron~\cite{hastie2009elements}. Even though using implicit features affected the performance of individual classification algorithms (positive in case of kNN and MNB, negative in case of RF), the final accuracy (the result of the best approach for each type of features) did not improve much compared to the TF-IDF~\cite{ramos2003using} baseline (0.86 vs 0.88 accuracy).

The described approaches are good at predicting general topics but are not applicable to defining tags that relate to specific algorithms. Since statements do not usually explicitly mention how to solve the problem, determining algorithmic tags requires essentially solving the problem, which is difficult even for humans. An additional complication is that the statements are often allegorical in phrasing.
Another drawback of statement analysis is that we usually have a relatively small number of problems, which may lead to overfitting --- a case when the model tries to memorize samples from the train set rather than generalize from them.

\subsection{Source Code Analysis}
One can analyze not only problem statements written in natural language, but also source code of the problem's solutions. 
Shalaby et al.~\cite{shalaby2017automatic} used a metric-based approach to source code vectorization. For each of the Codeforces solutions in their dataset, they calculated 30 different software metrics, such as the number of variables of a specific type (\eg int, string), lines of code, number of loops, number of nested loops, etc. The authors evaluated how well these features allow them to distinguish solutions to problems from different categories 
and how well they allow to predict a particular type of algorithm within a particular category. 
Their experiments showed that traditional classification algorithms and code metrics may be successfully applied to the task of classifying solutions.

Sudha et al.~\cite{sudha2017classification} applied CNN character-wise, \ie interpreted each solution as just a sequence of separate characters. The authors used a dataset of problems from Codeforces and trained a CNN-based model to classify solutions into four classes. 
They also proposed to combine the information from all submitted solutions by choosing the most popular class from the classification of individual solutions. 
In their experiments, they managed to distinguish problems between four categories with an accuracy of $0.61$.

One disadvantage of these approaches is that tags are usually assigned to a problem as whole, so there is no guarantee that all of the tags will be relevant to every solution, since there could be several different ways to solve a problem.

\newcommand{\predictorname}{TagPredictor}
\newcommand{\predictor}{\textsc{\predictorname}\xspace}

\section{Proposed Approach}\label{sec:approach}

This section describes the proposed approach. 
Our main idea is to merge results of the text-based and the code-based models. To achieve that, we first trained a Gated Graph Neural Network~\cite{Allamanis2018LearningTR} to predict tags from solutions, then we fine-tuned BERT~\cite{devlin2018bert} to predict tags from problem statements, and finally combined both models into an ensemble.

\begin{figure}[t]
    \includegraphics[width=\columnwidth]{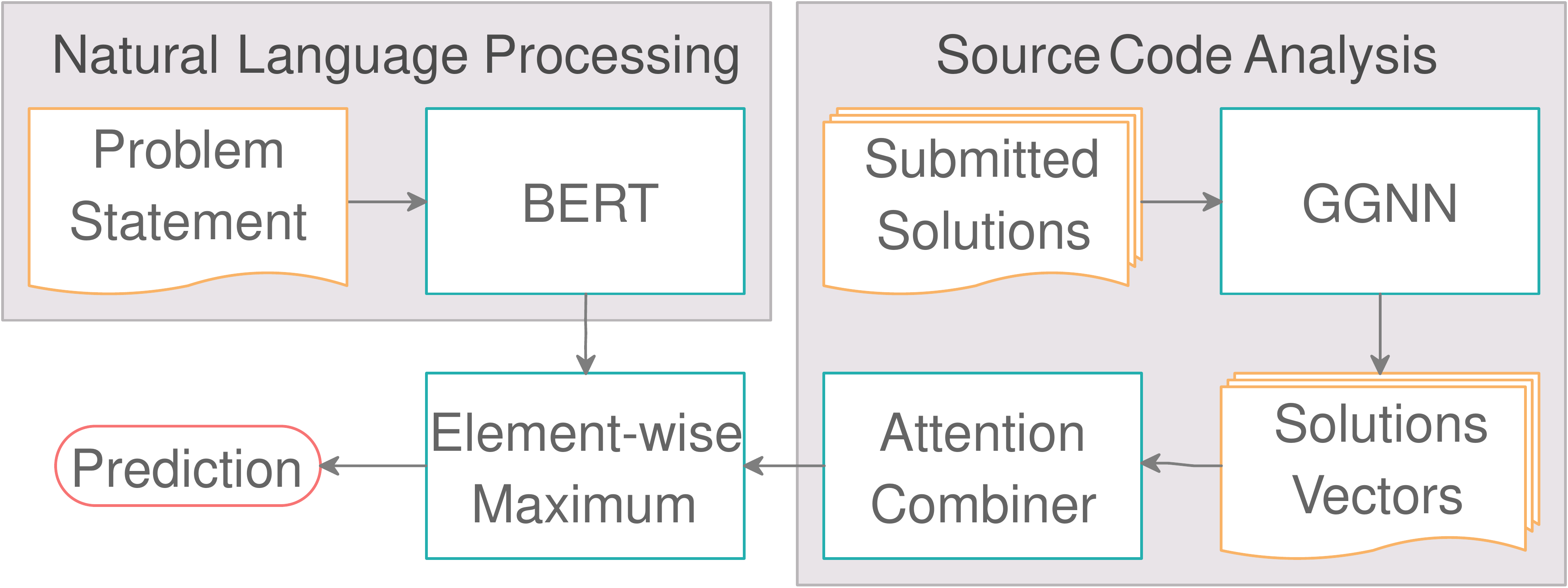}
	\caption{Pipeline of the proposed approach.}
	\label{fig:pipeline}
\vspace{-0.4cm}
\end{figure}

We will describe approaches that analyze problems or their solutions, which often lead us to the point where we have a vector representation of a solution or a problem, and we need to use it to predict probabilities of tags for this solution or a problem. To do this, we use a sequence of Dropout, Fully-Connected, and Sigmoid layers, which we will call \predictor from here on out. The Sigmoid layer maps the output for each tag to the $[0, 1]$ range, which may be interpreted as a probability of independent relevance. The Dropout layer is used for the regularization to deal with overfitting. 
The general pipeline is shown in \Cref{fig:pipeline}. Let us now describe each stage in more detail.

\subsubsection{Source code analysis}\label{sec:ggnn}

Graph Neural Network~\cite{zhou2018graph} is a deep learning method
that operates on graph-shaped data. It works as follows: we associate a numerical state vector with each node in the graph and then run several iterations of message exchange rounds between adjacent nodes. These messages update state vectors of the nodes depending on the node types, their current state, and types of the edges between nodes.
Gated Graph Neural Network (GGNN) is a special modification of GNN, which uses Gated Recurrent Unit~\cite{cho2014learning} to update state vectors. GGNN demonstrated good results in recent studies related to semantic code analysis~\cite{Allamanis2020TypilusNT, Allamanis2018LearningTR, li2015gated}, so we chose it to process solutions in our work. Our model is based on the version of GGNN introduced by Allamanis et al.~\cite{Allamanis2018LearningTR}.

Graph representation of code is based on the Abstract Syntax Tree (AST) representation augmented with additional edges. To build ASTs, we used Tree-sitter~\cite{tree-sitter}, the parser generator tool and an incremental parsing library. 
In our work, we used the same set of edge types as Allamanis et al.~\cite{Allamanis2018LearningTR}. To get the representation of the whole graph rather than of separate nodes, we added an additional sink node that has incoming edges of a special type from all other nodes in the graph. The final state vector of the sink node may be interpreted as a vector representation of the whole graph. We fed that vector to \predictor to get probabilities of individual tags for this particular solution. The model was trained via back propagation using the Adam~\cite{kingma2014adam} optimization algorithm. In our experiments, we started with hyperparameters described by Allamanis et al.~\cite{Allamanis2018LearningTR}, slightly varied them, and compared the results. Finally, we decided on the following model hyperparameters: the size of hidden node states and messages --- 128, the size of token embeddings --- 32, the size of variable type embedding --- 8, the number of message exchange rounds --- 5.

\subsubsection{Combining solutions}\label{sec:attention}
To aggregate the information we get for individual solutions and predict tags for a problem as a whole, we froze the weights of the GGNN model trained on individual solutions. Then, we applied the model to calculate vector representations for each submitted solution, combined them via the attention mechanism~\cite{bahdanau2014neural} into a single vector, and fed it to \predictor.
We fitted the parameters of the combining model with the Adam optimization algorithm. To prevent overfitting, during training, we used only up to 50 randomly sampled submitted solutions.

\subsubsection{Statement analysis}\label{sec:bert}

Bidirectional Encoder Representations from Transformers (BERT)~\cite{devlin2018bert} is a modern NLP method that has recently demonstrated state-of-the-art results on several NLP tasks~\cite{sun2019utilizing, liu2019text, peters2019tune}. One of the important ideas behind BERT is to pre-train the model on a large dataset once, and then just fine-tune it for every new task. It allows to achieve good performance even for small datasets due to the use of prior knowledge about the language. Thus, BERT looks like a suitable option for our task. 
The WordPieces~\cite{wu2016google} technique was applied to split tokens into segments, which helps to deal with out-of-vocabulary words. The sequence of tokens was extended by the special initial token, the final corresponding vector of which is used as text vector representation. We processed the resulting vector through \predictor to apply it to our task.

\subsubsection{Ensemble}\label{sec:ensemble}

At the last step, we combined GGNN and BERT into an ensemble for final predictions. 
These models gave us two probability vectors for each object. We could train a new model over that representation, but since these models were originally trained on the same dataset, they give much better results on the train set, so such data representation is inconsistent. Therefore, we calculated the element-wise maximum of probability vectors as the natural pooling method.

\section{Dataset}\label{sec:dataset}

In our study, we used a dataset of problems from the Codeforces competitive programming platform. 
The dataset contains 1,065 contests, 5,981 problems, and 14,257,576 solutions.
For each problem, we have a set of manually assigned relevant tags, a statement, and a set of submitted solutions that passed all the existing tests. In this work, we only analyzed solutions written in C++, since code-based approaches are language-dependent and C++ is by far the most popular language for competitive programming. The number of solutions may vary from thousands for easy and popular problems to only a few for the hardest ones. 
There are 35 different tags in the dataset, but 5 of them are too rare, so we skip them.

\subsection{Deduplication}\label{sec:deduplication}
Solutions in the dataset are not unique. We used the SHA‑256~\cite{gilbert2003security} algorithm to calculate the hash code of every solution, then groups of solutions with the same hash codes were removed, except for the earliest solution in each group.

However, not only solutions can be duplicated, but problem statements may be repeated as well. Nearly identical statements often correspond to the same problem with different difficulty requirements (limit of input size, execution time, and memory). The simplified version of a problem may accept dummy solutions using \textit{brute force}, while a more advanced version of the same problem may require sophisticated solutions involving the use of advanced algorithms. To avoid adding extra irrelevant labels to individual solutions, we decided to neither remove nor combine such problems. We simply made sure that none of the problems in the test set were used at the training stage or during the hyperparameter optimization. We labeled statements as nearly identical if the similarity of their token sets was at least 0.9 according to Jaccard index~\cite{gower2014similarity}. All in all, on this stage, we removed 18 problems and 2.5\% of solutions.

\subsection{Data Splitting}

The first contest in our dataset was held in 2010, while the last contest took place in 2019. A lot of things changed during those years (\eg new constructs were introduced in the programming language, the platform's audience was growing), so the dataset is not homogeneous and that might have an effect on the model's performance. 
An illustration of this may be found in the study by Shalaby~et al.~\cite{shalaby2017automatic}. The authors compared the results of the same approach on the randomly split and on the chronologically split dataset, and observed a significant drop of performance metrics in the latter case. 

Thus, to get reliable evaluation results, we split our dataset of contests chronologically.
To guarantee that the splitting is correct, we removed all solutions from the train dataset that were submitted later than the first submission in the validation dataset. The same procedure was applied to the validation and test datasets. Finally, to reduce the data imbalance, we limit the maximum number of solutions per problem to 1,000 and randomly reduce the number of solutions to 1,000 if a problem has more than a 1,000 solutions. \Cref{tab:dataset_split} describes all the dataset parts after deduplication and splitting.

\begin{table}
  \caption{Final dataset after deduplication and subsampling.}
  \label{tab:dataset_split}
  \begin{center}
    \begin{tabular}{ l c c c }
    \toprule
    \multicolumn{1}{c}{\textbf{Dataset}} & \textbf{Contests} & \textbf{Problems} & \textbf{Solutions}\\
    \midrule
    Train & 716 & 3,837 & 2,493,745 \\
    Validation & 169 & 1,020 & 664,578 \\
    Test & 159 & 1,046& 734,612  \\
    \midrule
    Total &	1,044   &  5,903  & 3,892,935 \\
    \bottomrule
        \end{tabular}
        \end{center}
\vspace{-0.4cm}
\end{table}

\subsection{Preprocessing}\label{sec:preprocessing}
The analysis of source code in C++ is a complicated task due to the rich grammar of the language. To slightly simplify it, we removed all \textit{include} macros to avoid including header files from the standard library, ran the compiler's preprocessor to substitute user-defined macros, and removed comments. 

Problem statements require preprocessing too. Initially, statements are stored in the LaTeX format, which is not suitable for NLP analysis straight away. We used the Pandoc tool~\cite{dominici2014overview} to convert statements into plain text and then applied the pre-trained WordPieces~\cite{wu2016google} tokenizer to get text representation that BERT~\cite{devlin2018bert} receives as input. 
\newcommand{\scoretable}[3]{
    \begin{table}
        \small
        \caption{#3}
        \vspace{-0.2cm}
        \label{tab:#1}
        \csvreader[
            centered tabular= l c c c c,
            table head=\toprule \multicolumn{1}{c}{\textbf{Approach}} & \textbf{PR-AUC} & \textbf{F1} & \textbf{P} & \textbf{R}\\\midrule,
            late after line = \ifcsvstrcmp{\sepline}{True}{\\\midrule}{\\},
            late after last line=\\\bottomrule]%
            {#2}{approach=\approach,paper=\paper,prauc=\prauc,fscore=\fscore,precision=\precision,recall=\recall,sepline=\sepline}%
            {\approach\ifcsvstrcmp{\paper}{}{}{~\cite{\paper}} & \prauc & \fscore & \precision & \recall}%
    \vspace{-0.6cm}
    \end{table}
}

\section{Evaluation}\label{sec:evaluation}

This section provides a comparison of different sources of information and different approaches to tag prediction.
We discuss evaluation metrics, overview the baseline approaches, describe evaluation experiments, and present their results. 

\subsection{Evaluation Metrics}

Precision and Recall~\cite{powers2011evaluation} are the universally applied classification metrics, which measure the prediction quality from both sides: correctness and completeness. However, since classifiers usually provide the probability of an object belonging to a certain class, we can affect these metrics by changing the classification threshold. Increasing the threshold tends to increase Precision and decrease Recall (because we choose less but with greater confidence), and vice versa. Choosing a threshold value is always a balance between these metrics, and its specific choice depends on the conditions for applying the model. For these reasons, these metrics are not usually used to directly compare the quality of models. 

F1 score~\cite{powers2011evaluation}, the harmonic mean of Precision and Recall, is more suitable for comparison, but it still depends on the selected threshold value. 
One way to avoid this problem is to analyze the Precision-Recall Curve, which is a plot of Precision from Recall as classification threshold is changed. The Area Under Precision-Recall Curve (PR-AUC)~\cite{davis2006relationship} is a popular classification metric, robust to highly skewed datasets.

We chose PR-AUC as the main metric in our experiments. We also provide F1 score, Precision, and Recall, threshold values for which are fitted on the validation set maximizing the F1 score.

\subsection{Text Analysis}

In the first experiment, we evaluated the usefulness of problem statements for predicting tags. For this purpose, we took the statements of all problems from the train set (see \Cref{sec:dataset}) and trained several textual-based models to predict their tags. The full list of applied models is the following:

\begin{enumerate}
    \item Bora et al.~\cite{Bora2016PredictingAA}: LSTM over one-hot encoding (OHE), LSTM over word2vec, Logistic Regression (LR) over Bag-of-Words (BoW). Following the original paper, we did not apply any NLP preprocessing techniques (\eg stemming, removal of stop-words). LSTM models were pre-trained on the task of predicting difficulty of the problems. There is no strict difficulty categories on Codeforces though, so we used contest divisions (some contests targets beginners, other --- more advanced participants) instead of them. 
    \item Athavale et al.~\cite{shrivastava2019predicting}: CNN over trainable word embeddings (TWE + CNN), CNN over pre-trained GloVe~\cite{pennington2014glove} word embeddings (GloVe + CNN) and ensemble of CNN models. The authors published their source code, so we used their implementation.
    \item Iancu et al.~\cite{iancu2019multi}: LSTM over one-hot encoding, LSTM over word2vec, Decision Tree over TF-IDF. We used Natural Language Toolkit~\cite{nltk} to remove stop words.
    \item Our work: the BERT~\cite{devlin2018bert} model. We used a pre-trained model and fine-tuned it on our task. For more details, see \Cref{sec:bert}.
\end{enumerate}

The results of these experiments are presented in \Cref{tab:scores_text}. It could be noted that traditional classification algorithms such as Decision Tree show quite good results, which is consistent with the findings presented by Bora et al.~\cite{Bora2016PredictingAA} The possible reason is that they have a relatively small amount of parameters so they are more resistant to overfitting than deep learning models. All LSTM based models performed a bit worse. The common problem of LSTM in classification tasks is gradient vanishing~\cite{pascanu2013difficulty}, the negative effect of back-propagation in deep networks. In contrast, the CNN based model does not suffer from that problem. The ensemble of them is more robust, so it is not surprising that it shows the best results of $0.278$ PR-AUC. The BERT model was only slightly worse by PR-AUC ($0.273$ vs $0.278$), but better by F1 ($0.308$ vs $0.289$). The advantage of BERT is its pre-training on the large dataset. 

This experiment demonstrates that our model is capable of performing on par with other excising models when predicting tags from problem statements. 

\scoretable{scores_text}{tables/scores-text-based.csv}%
    {Evaluation of statement-based approaches for predicting the tags of problems. The bottom line represents this work.}

\subsection{Source Code Analysis}
Another source of information about the problem is the list of submitted solutions.  We perform our experiments on its usefulness in two stages. In the first stage, the task was to predict tags for individual solutions from the dataset. Unfortunately, we do not have tags for each particular solution, so we assumed the tags of each problem to fit all of its solutions. In the second stage, the task was to predict the tags for problems given all of their submitted solutions. The following models were compared:

\begin{enumerate}
    \item Shalaby et al.~\cite{shalaby2017automatic}: AdaBoost, Random Forest (RF), and Support Vector Machine (SVM) from sklearn~\cite{sklearn} over hand-crafted set of software metrics.
    \item Sudha et al.~\cite{sudha2017classification}: character-wise CNN. 
    \item Our work: GGNN to analyze individual solutions, GGNN + Attention to predict tags to problems. For more details see \Cref{sec:ggnn}.
\end{enumerate}
 
The results of the first stage of the experiment are shown in \Cref{tab:scores_code_single}. Similarly to the study of Bora et al.~\cite{Bora2016PredictingAA}, Random Forest reached $0.254$ PR-AUC, the best result between all metric-based approaches.
Character-wise CNN is a more flexible approach than metrics analysis, but it still performed worse than Random Forest. It also completely ignores the information about the code structure.
Because of that, GGNN model drastically outperforms all of the baselines in a single solution classification task. 

\scoretable{scores_code_single}{tables/scores-code-based-single.csv}%
    {Evaluation of code-based approaches for predicting the tags of solutions. The bottom line represents this work.}

In the second stage of the experiment, we compared the quality of predicting tags for a problem given all of its solutions. The similar task was investigated in the study of Sudha et al.~\cite{sudha2017classification}. They propose to train a model on the task of classifying individual solutions (like in the first stage of our evaluation), predict a class for every solution submitted to the problem, and claim the most popular result as a class for the problem. We used that idea to apply metric-based approaches and character-wise CNN trained in the first stage to predicting tags for problems. In contrast to the original study, we work with a multi-label task, so we apply that approach for each individual tag. 
We proposed another way to aggregate the solutions. The GGNN model from the first stage was used to vectorize all the solutions, and then we train an attention-based model that uses the obtained vector representation of solutions to predict tags for the entire problem.

The results of the second stage are presented in \Cref{tab:scores_code}. As was expected, the results of classifying the whole problems are better than the results of classifying just individual solutions, because now we can take into account different ways of solving the problem and reduce the impact of noisy solutions. Since we employed the models that were trained during the first stage of the experiment, it is not surprising that these results correlate with the results presented in~\Cref{tab:scores_code_single}. The metric values of all the approaches increased, but our approach still demonstrated the best PR-AUC score.

\scoretable{scores_code}{tables/scores-code-based.csv}%
    {Results of code-based approaches for predicting the tags of problems. The bottom line represents this work.}

\subsection{Tags Prediction}\label{sec:analysis}

\begin{figure}
    \vspace{0.3cm}
    \centering
    \includegraphics[width=\columnwidth]{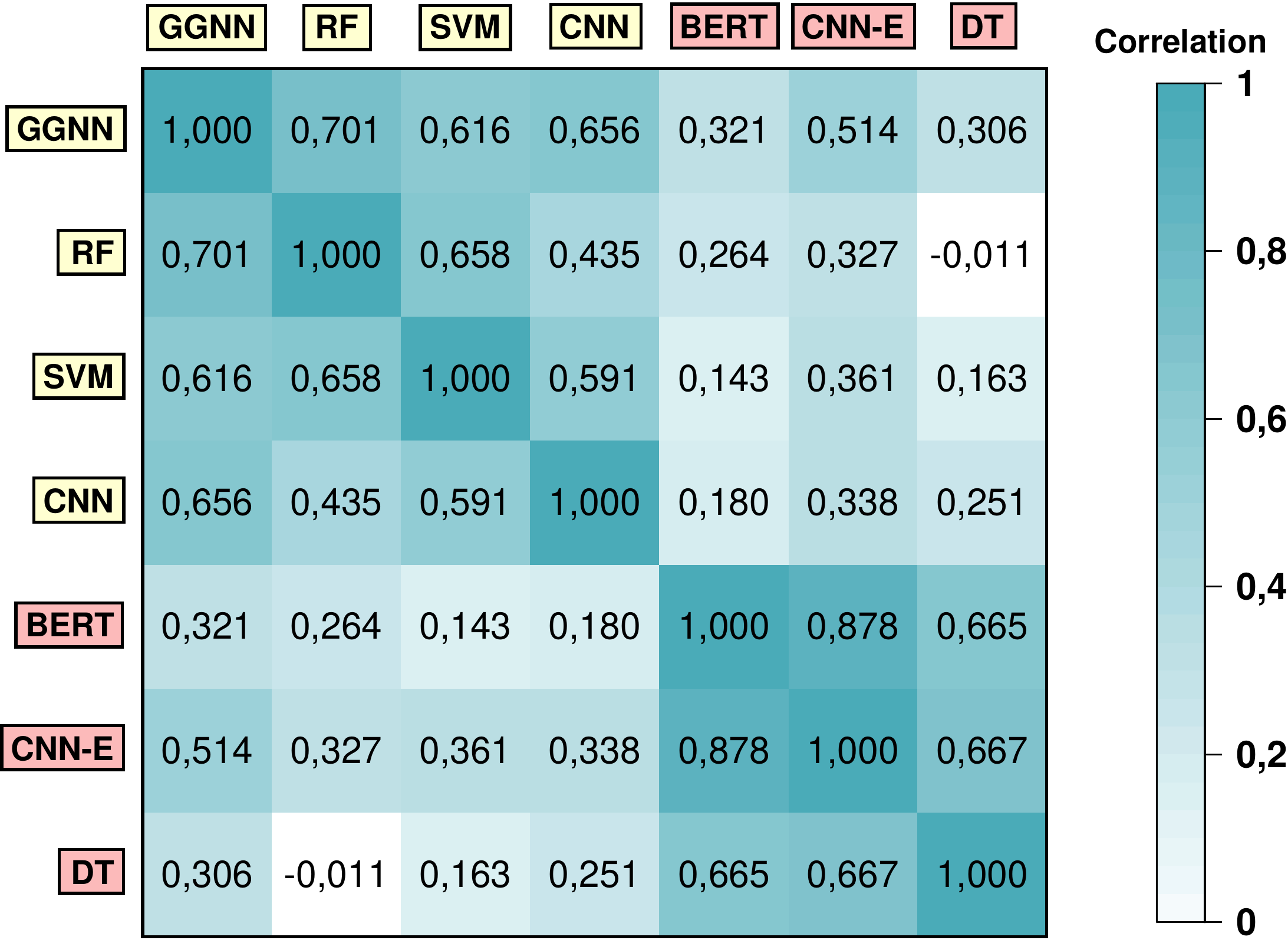}
    \caption{Correlations of lists of PR-AUC values for several approaches.}
	\label{fig:correlations}
	\vspace{-0.2cm}
\end{figure}

In the previous sections, we explored two points of view onto the problem of predicting tags. At the first glance, it seems that code-based models are certainly better than statement-based ones. However, this impression appears because the tables show only the averaged metrics between all tags. The real situation is more complicated. To illustrate this idea, we selected several models and calculated lists of their PR-AUC values for individual tags and then measured the Pearson correlation coefficient~\cite{benesty2009pearson} of these lists. The high coefficient means that values of lists change similarly, \ie if one model predicts a separate tag better than average, then the second one is expected to predict that tag relatively better as well. The results are shown in \Cref{fig:correlations}. Code-based models are highlighted in yellow, text-based models --- in red. As the figure demonstrates, correlations between models of the same type are higher than between types, which proves that models of different types tend to be better for different sets of tags. That is why it is important to use both types of information. 

As descried in \Cref{sec:ensemble}, we combine GGNN and BERT models into an ensemble as representatives of different types of approaches. The comparison of the ensemble with other approaches is shown in \Cref{tab:scores_final}. Our model outperformed all other models, including GGNN and BERT individually. 

\scoretable{scores_final}{tables/scores-final.csv}{Final results. The bottom lines represent this work.}

\subsection{Motivating Example}
The model gives us an opportunity to check the correctness of existing labels by examining problems, where the model is very confident in its answer, but it does not match the current label. For example, our model suggested to add tags \textit{strings} and \textit{hashing} to the problem \textit{1073/G}~\cite{problem}. The problem does not have these tags at the moment, but it is definitely about processing strings, and hashing may be applied to find the longest common prefix like in the submission \textit{№44992266}~\cite{submission}.
\section{Threats to Validity}\label{sec:threats-to-validity}

\subsubsection*{Internal Validity}

Only one of the related papers has a publicly available implementation, so we had to reproduce other approaches ourselves, and we could have made errors. However, because most of the employed models are typical to the research area and the most important implementation details were described in corresponding papers, we believe that the threat does not significantly impact our results. Another possible problem is that the dataset we used have been labeled by many different people during the course of almost ten years, so some tags may be noisy or inconsistent. However, since only the authors of the problems and highly experienced users who solved them were allowed to provide tags, we expect them to be mostly correct. 

\subsubsection*{External Validity}
We used the dataset from a single competitive programming platform, so the results may differ for other problem databases. However, we expect the dataset to still be representative, since Codeforces is one of the most popular platforms and we did not perform any specific problem filtering.
We also cannot guarantee the same prediction quality for solutions in other programming languages. However, C++ is the most popular language for competitive programming, so we believe that our results are still valuable. 

Overall, while these threats are important to note, we believe that we mitigated them well and they do not invalidate the results of our research.
\section{Conclusion}\label{sec:conclusion}

In this paper, we present a novel method for the automatic tag prediction for competitive programming problems. It is based on the idea of combining source code analysis and natural language processing. We trained a Gated Graph Neural Network with the attention mechanism to analyze the submitted solutions, fine-tuned the BERT model to process problems statements, and combined these models into an ensemble.

We implemented a wide range of existing approaches from five related papers and evaluated all of them on a dataset of problems from Codeforces. The analysis of the correlation of the models' predictions demonstrates that text-based models and code-based models are good at predicting different sets of tags, therefore, combining them should increase the overall quality of predictions.
Indeed, our model outperforms the best text-based model by $0.264$ and the best code-based model by $0.175$ of the PR-AUC metric. The source code of the project and the trained models are available online~\cite{rp}.

Overall, while it is too early to say that manual labelling of problems in competitive programming is no longer necessary, our model can significantly simplify this procedure, as well as find errors in existing label sets. We believe that our study can provide benefits for both developers of competitive programming platforms and competitors, and hope that it will be useful for other researchers in this area.

\bibliographystyle{ieeetran}
\balance
\bibliography{references}

\begin{thebibliography}{10}
\providecommand{\url}[1]{#1}
\csname url@samestyle\endcsname
\providecommand{\newblock}{\relax}
\providecommand{\bibinfo}[2]{#2}
\providecommand{\BIBentrySTDinterwordspacing}{\spaceskip=0pt\relax}
\providecommand{\BIBentryALTinterwordstretchfactor}{4}
\providecommand{\BIBentryALTinterwordspacing}{\spaceskip=\fontdimen2\font plus
\BIBentryALTinterwordstretchfactor\fontdimen3\font minus
  \fontdimen4\font\relax}
\providecommand{\BIBforeignlanguage}[2]{{%
\expandafter\ifx\csname l@#1\endcsname\relax
\typeout{** WARNING: IEEEtran.bst: No hyphenation pattern has been}%
\typeout{** loaded for the language `#1'. Using the pattern for}%
\typeout{** the default language instead.}%
\else
\language=\csname l@#1\endcsname
\fi
#2}}
\providecommand{\BIBdecl}{\relax}
\BIBdecl

\bibitem{bloomfield2016programming}
A.~Bloomfield and B.~Sotomayor, ``A programming contest strategy guide,'' in
  \emph{Proceedings of the 47th ACM technical symposium on computing science
  education}, 2016, pp. 609--614.

\bibitem{codeforces}
\BIBentryALTinterwordspacing
CodeForces. (accessed: 01.01.2023) A competitive programming platform.
  [Online]. Available: \url{https://codeforces.com/}
\BIBentrySTDinterwordspacing

\bibitem{codechef}
\BIBentryALTinterwordspacing
CodeChef. (accessed: 01.01.2023) A competitive programming platform. [Online].
  Available: \url{https://www.codechef.com/}
\BIBentrySTDinterwordspacing

\bibitem{Bora2016PredictingAA}
A.~Bora and A.~Sinha, ``Predicting algorithmic approach for programming
  problems from natural language problem description,'' 2016.

\bibitem{shrivastava2019predicting}
V.~Athavale, A.~Naik, R.~Vanjape, and M.~Shrivastava, ``Predicting algorithm
  classes for programming word problems,'' in \emph{Proceedings of the 5th
  Workshop on Noisy User-generated Text (W-NUT 2019)}, 2019, pp. 84--93.

\bibitem{iancu2019multi}
B.~Iancu, G.~Mazzola, K.~Psarakis, and P.~Soilis, ``Multi-label classification
  for automatic tag prediction in the context of programming challenges,''
  \emph{arXiv preprint arXiv:1911.12224}, 2019.

\bibitem{shalaby2017automatic}
M.~Shalaby, T.~Mehrez, A.~El~Mougy, K.~Abdulnasser, and A.~Al-Safty,
  ``Automatic algorithm recognition of source-code using machine learning,'' in
  \emph{2017 16th IEEE International Conference on Machine Learning and
  Applications (ICMLA)}, 2017, pp. 170--177.

\bibitem{sudha2017classification}
S.~Sudha, A.~A. Kumar, M.~M. Nagappan, and R.~Suresh, ``Classification and
  recommendation of competitive programming problems using {CNN},'' in
  \emph{International Conference on Intelligent Information Technologies},
  2017, pp. 262--272.

\bibitem{intisar2019classification}
C.~M. Intisar, Y.~Watanobe, M.~Poudel, and S.~Bhalla, ``Classification of
  programming problems based on topic modeling,'' in \emph{Proceedings of the
  2019 7th International Conference on Information and Education Technology},
  2019, pp. 275--283.

\bibitem{rp}
\BIBentryALTinterwordspacing
R.~Package. (accessed: 01.01.2023) Predicting tags for programming tasks.
  [Online]. Available:
  \url{https://github.com/JetBrains-Research/tag-prediction}
\BIBentrySTDinterwordspacing

\bibitem{hochreiter1997long}
S.~Hochreiter and J.~Schmidhuber, ``Long short-term memory,'' \emph{Neural
  computation}, vol.~9, no.~8, pp. 1735--1780, 1997.

\bibitem{mikolov2013distributed}
T.~Mikolov, I.~Sutskever, K.~Chen, G.~S. Corrado, and J.~Dean, ``Distributed
  representations of words and phrases and their compositionality,'' in
  \emph{Advances in neural information processing systems}, 2013, pp.
  3111--3119.

\bibitem{ho1995random}
T.~K. Ho, ``Random decision forests,'' in \emph{Proceedings of 3rd
  international conference on document analysis and recognition}, vol.~1, 1995,
  pp. 278--282.

\bibitem{Kim2014ConvolutionalNN}
Y.~Kim, ``Convolutional neural networks for sentence classification,'' in
  \emph{Proceedings of the 2014 Conference on Empirical Methods in Natural
  Language Processing (EMNLP)}, 2014, pp. 1746--1751.

\bibitem{topcoder}
\BIBentryALTinterwordspacing
TopCoder. (accessed: 01.01.2023) A competitive programming platform. [Online].
  Available: \url{https://www.topcoder.com/}
\BIBentrySTDinterwordspacing

\bibitem{Quoc2014documents}
Q.~V. Le and T.~Mikolov, ``Distributed representations of sentences and
  documents,'' in \emph{International Conference on Machine Learning}, 2014.

\bibitem{hamming1950}
R.~W. {Hamming}, ``Error detecting and error correcting codes,'' \emph{The Bell
  System Technical Journal}, vol.~29, no.~2, pp. 147--160, 1950.

\bibitem{Blei2003LatentDA}
D.~M. Blei, A.~Y. Ng, and M.~I. Jordan, ``Latent dirichlet allocation,''
  \emph{J. Mach. Learn. Res.}, vol.~3, pp. 993--1022, 2003.

\bibitem{Lee2000AlgorithmsFN}
D.~D. Lee and H.~S. Seung, ``Algorithms for non-negative matrix
  factorization,'' in \emph{Advances in neural information processing systems},
  2001, pp. 556--562.

\bibitem{larose2014}
D.~T. Larose and C.~D. Larose, \emph{Discovering knowledge in data: an
  introduction to data mining}, 2014.

\bibitem{rennie2003tackling}
J.~D. Rennie, L.~Shih, J.~Teevan, and D.~R. Karger, ``Tackling the poor
  assumptions of {Naive Bayes} text classifiers,'' in \emph{Proceedings of the
  20th international conference on machine learning (ICML-03)}, 2003, pp.
  616--623.

\bibitem{hastie2009elements}
T.~Hastie, R.~Tibshirani, and J.~Friedman, \emph{The elements of statistical
  learning: data mining, inference, and prediction}, 2009.

\bibitem{ramos2003using}
J.~Ramos \emph{et~al.}, ``Using {TF-IDF} to determine word relevance in
  document queries,'' in \emph{Proceedings of the first instructional
  conference on machine learning}, vol. 242, 2003, pp. 133--142.

\bibitem{Allamanis2018LearningTR}
M.~Allamanis, M.~Brockschmidt, and M.~Khademi, ``Learning to represent programs
  with graphs,'' \emph{ArXiv}, vol. abs/1711.00740, 2018.

\bibitem{devlin2018bert}
J.~Devlin, M.-W. Chang, K.~Lee, and K.~Toutanova, ``{BERT}: Pre-training of
  deep bidirectional transformers for language understanding,'' in
  \emph{Proceedings of NAACL-HLT}, 2019, pp. 4171--4186.

\bibitem{zhou2018graph}
J.~Zhou, G.~Cui, S.~Hu, Z.~Zhang, C.~Yang, Z.~Liu, L.~Wang, C.~Li, and M.~Sun,
  ``Graph neural networks: A review of methods and applications,'' \emph{AI
  Open}, vol.~1, pp. 57--81, 2020.

\bibitem{cho2014learning}
K.~Cho, B.~van Merrienboer, {\c{C}}.~G{\"u}l{\c{c}}ehre, D.~Bahdanau,
  F.~Bougares, H.~Schwenk, and Y.~Bengio, ``Learning phrase representations
  using {RNN} encoder-decoder for statistical machine translation,'' in
  \emph{EMNLP}, 2014.

\bibitem{Allamanis2020TypilusNT}
M.~Allamanis, E.~T. Barr, S.~Ducousso, and Z.~Gao, ``Typilus: neural type
  hints,'' in \emph{Proceedings of the 41st ACM SIGPLAN Conference on
  Programming Language Design and Implementation}, 2020, pp. 91--105.

\bibitem{li2015gated}
Y.~Li, R.~Zemel, M.~Brockschmidt, and D.~Tarlow, ``Gated graph sequence neural
  networks,'' in \emph{Proceedings of ICLR'16}, 2016.

\bibitem{tree-sitter}
\BIBentryALTinterwordspacing
Tree-sitter. (accessed: 01.01.2023) Parser generator tool. [Online]. Available:
  \url{https://tree-sitter.github.io/}
\BIBentrySTDinterwordspacing

\bibitem{kingma2014adam}
D.~P. Kingma and J.~Ba, ``Adam: A method for stochastic optimization,''
  \emph{arXiv preprint arXiv:1412.6980}, 2014.

\bibitem{bahdanau2014neural}
D.~Bahdanau, K.~H. Cho, and Y.~Bengio, ``Neural machine translation by jointly
  learning to align and translate,'' in \emph{3rd International Conference on
  Learning Representations, ICLR 2015}, 2015.

\bibitem{sun2019utilizing}
C.~Sun, L.~Huang, and X.~Qiu, ``Utilizing {BERT} for aspect-based sentiment
  analysis via constructing auxiliary sentence,'' in \emph{Proceedings of
  NAACL-HLT}, 2019, pp. 380--385.

\bibitem{liu2019text}
Y.~Liu and M.~Lapata, ``Text summarization with pretrained encoders,'' in
  \emph{Proceedings of the 2019 Conference on Empirical Methods in Natural
  Language Processing and the 9th International Joint Conference on Natural
  Language Processing (EMNLP-IJCNLP)}, 2019, pp. 3730--3740.

\bibitem{peters2019tune}
M.~E. Peters, S.~Ruder, and N.~A. Smith, ``To tune or not to tune? {Adapting}
  pretrained representations to diverse tasks,'' in \emph{Proceedings of the
  4th Workshop on Representation Learning for NLP (RepL4NLP-2019)}, 2019, pp.
  7--14.

\bibitem{wu2016google}
Y.~Wu, M.~Schuster, Z.~Chen, Q.~V. Le, M.~Norouzi, W.~Macherey, M.~Krikun,
  Y.~Cao, Q.~Gao, K.~Macherey \emph{et~al.}, ``Google's neural machine
  translation system: Bridging the gap between human and machine translation,''
  \emph{arXiv preprint arXiv:1609.08144}, 2016.

\bibitem{gilbert2003security}
H.~Gilbert and H.~Handschuh, ``Security analysis of {SHA-256} and sisters,'' in
  \emph{International workshop on selected areas in cryptography}, 2003, pp.
  175--193.

\bibitem{gower2014similarity}
J.~C. Gower and M.~J. Warrens, ``Similarity, dissimilarity, and distance,
  measures of,'' \emph{Wiley StatsRef: Statistics Reference Online}, pp. 1--11,
  2014.

\bibitem{dominici2014overview}
M.~Dominici, ``An overview of {Pandoc},'' \emph{TUGboat}, vol.~35, no.~1, pp.
  44--50, 2014.

\bibitem{powers2011evaluation}
D.~M. Powers, ``Evaluation: from precision, recall and {F-measure} to {ROC},
  informedness, markedness and correlation,'' 2011.

\bibitem{davis2006relationship}
J.~Davis and M.~Goadrich, ``The relationship between precision-recall and {ROC}
  curves,'' in \emph{Proceedings of the 23rd international conference on
  Machine learning}, 2006, pp. 233--240.

\bibitem{pennington2014glove}
J.~Pennington, R.~Socher, and C.~D. Manning, ``Glove: Global vectors for word
  representation,'' in \emph{Proceedings of the 2014 conference on empirical
  methods in natural language processing (EMNLP)}, 2014, pp. 1532--1543.

\bibitem{nltk}
\BIBentryALTinterwordspacing
NLTK. (accessed: 01.01.2023) Natural language toolkit. [Online]. Available:
  \url{https://www.nltk.org/}
\BIBentrySTDinterwordspacing

\bibitem{pascanu2013difficulty}
R.~Pascanu, T.~Mikolov, and Y.~Bengio, ``On the difficulty of training
  recurrent neural networks,'' in \emph{International conference on machine
  learning}, 2013, pp. 1310--1318.

\bibitem{sklearn}
\BIBentryALTinterwordspacing
SKLearn. (accessed: 01.01.2023) scikit-learn. [Online]. Available:
  \url{https://scikit-learn.org/}
\BIBentrySTDinterwordspacing

\bibitem{benesty2009pearson}
J.~Benesty, J.~Chen, Y.~Huang, and I.~Cohen, ``Pearson correlation
  coefficient,'' in \emph{Noise reduction in speech processing}, 2009, pp.
  1--4.

\bibitem{problem}
\BIBentryALTinterwordspacing
P.~1073/G. (accessed: 01.01.2023) A problem about strings. [Online]. Available:
  \url{https://codeforces.com/contest/1073/problem/G}
\BIBentrySTDinterwordspacing

\bibitem{submission}
\BIBentryALTinterwordspacing
S.~№44992266. (accessed: 01.01.2023) Submission №44992266. [Online].
  Available: \url{https://codeforces.com/contest/1073/submission/44992266}
\BIBentrySTDinterwordspacing

\end{thebibliography}

\end{document}